\documentclass[]{pasj01}

\Received{$\langle$reception date$\rangle$}
\Accepted{$\langle$acception date$\rangle$}
\Published{$\langle$publication date$\rangle$}

\usepackage{graphicx}    
\usepackage{tabulary}
\usepackage{booktabs}
\usepackage{footnote}
\usepackage{hyperref}
\usepackage{natbib}
\usepackage[dvipsnames]{xcolor}
\usepackage[symbol]{footmisc}
\bibpunct{(}{)}{;}{a}{}{,}
\renewcommand{\thefootnote}{\fnsymbol{footnote}}

\newcommand{\B}[0]{\mathsf{B}}
\newcommand{\W}[0]{W}
\newcommand{\F}[0]{\mathcal{F}}
\newcommand{\WT}[0]{\mathcal{W}}
\newcommand{\I}[0]{\mathsf{I}}

\DeclareRobustCommand{\VAN}[3]{#2}
\let\VANthebibliography\thebibliography
\def\thebibliography{\DeclareRobustCommand{\VAN}[3]{##3}\VANthebibliography}

\begin{document}

\title{Wavelets and sparsity for Faraday tomography}

\author{
Suchetha \textsc{Cooray},\altaffilmark{1,}$^{*, }$\footnotemark[2]
Tsutomu T. Takeuchi,$^{1,2}$
Shinsuke Ideguchi,$^{3}$
Takuya Akahori,$^{4, 5}$
Yoshimitsu Miyashita,$^{6}$
and Keitaro Takahashi,$^{6, 7, 8}$
}

\altaffiltext{1}{Division of Particle and Astrophysical Science, Nagoya University, Furo-cho, Chikusa-ku, Nagoya 464–8602, Japan}
\altaffiltext{2}{The Research Center for Statistical Machine Learning, The Institute of Statistical Mathematics, 10-3 Midori-cho, Tachikawa, Tokyo 190-8562, Japan}
\altaffiltext{3}{Department of Astrophysics/IMAPP, Radboud University Nijmegen, PO Box 9010, NL-6500 GL Nijmegen, the Netherlands}
\altaffiltext{4}{Mizusawa VLBI Observatory, National Astronomical Observatory of Japan, 2-21-1 Osawa, Mitaka, Tokyo 181-8588, Japan}
\altaffiltext{5}{SKA Organization, Jodrell Bank, Lower Withington, Macclesfield, SK11 9DL, UK}
\altaffiltext{6}{Kumamoto University, 2-39-1, Kurokami, Kumamoto 860-8555, Japan}
\altaffiltext{7}{International Research Organization for Advanced Science and Technology, Kumamoto University, Japan}
\altaffiltext{8}{National Astronomical Observatory of Japan, 2-21-1 Osawa, Mitaka, Tokyo 181-8588, Japan}


\email{cooray@nagoya-u.jp}

\KeyWords{magnetic fields -- polarization -- techniques: polarimetric -- techniques: interferometric -- methods: data analysis}

\maketitle

\begin{abstract}
    Faraday tomography through broadband polarimetry can provide crucial information on magnetized astronomical objects, such as quasars, galaxies, or galaxy clusters. However, the limited wavelength coverage of the instruments requires that we solve an ill-posed inverse problem when we want to obtain the Faraday dispersion function (FDF), a tomographic distribution of the magnetoionic media along the line of sight. This paper explores the use of wavelet transforms and the sparsity of the transformed FDFs in the form of wavelet shrinkage (WS) for finding better solutions to the inverse problem. We recently proposed the \textit{Constraining and Restoring iterative Algorithm for Faraday Tomography} (CRAFT; Cooray et al. 2021), a new flexible algorithm that showed significant improvements over the popular methods such as Rotation Measure Synthesis. In this work, we introduce CRAFT+WS, a new version of CRAFT incorporating the ideas of wavelets and sparsity. CRAFT+WS exhibit significant improvements over the original CRAFT when tested for a complex FDF of realistic Galactic model. Reconstructions of FDFs demonstrate super-resolution in Faraday depth, uncovering previously unseen Faraday complexities in observations. The proposed approach will be necessary for effective cosmic magnetism studies using the Square Kilometre Array and its precursors. The code is made publicly available\footnotemark[3].

\end{abstract}
    
\footnotetext[2]{Research Fellow of the Japan Society for the Promotion of Science (DC1)}
\footnotetext[3]{{\url{https://github.com/suchethac/craft}}}


\section{Introduction} \label{sec:introduction}

    Understanding cosmic magnetism from the scale of interstellar gas to galaxy clusters is crucial in understanding their astrophysical processes \citep[e.g.,][]{Gaensler_2004, Beck_2009, Johnston-Hollitt_2015, Akahori_2016, Akahori_2018a}. One of the main techniques to obtain the magnetic information of these objects is through observations of polarized radio emission. When polarized light passes through magnetoionic media, the radiation experiences frequency-dependent faraday rotation. The frequency-dependent rotation can be dissected along the line-of-sight (LOS) with multichannel radio polarimetry to trace the magnetic structures \citep{Kronberg_1982, Kolatt_1998, Stasyszyn_2010, Akahori_2014}. Initially popularized by the seminal paper of \citet{Burn_1966}, this process is called \textit{Faraday tomography} and is now a popular technique to analyze polarization data.
    
    In Faraday tomography, the linear polarization components (Stokes $Q$ and Stokes $U$) gives us the complex linear polarization spectrum $P=Q+iU$. $P$ is then used to obtain the Faraday dispersion function $F(\phi)$ (FDF) through the relation,
    \begin{equation} \label{eq:FT}
        P\left(\lambda^{2}\right)=\int_{0}^{\infty} \varepsilon(r) e^{2 i \chi\left(r, \lambda^{2}\right)} \mathrm{d} r =\int_{-\infty}^{\infty} F(\phi) e^{2 i \phi \lambda^{2}} \mathrm{d} \phi ,
    \end{equation}
    where $\lambda$ is the wavelength of the polarized emission, $r$ is the physical distance from the observer, $\varepsilon$ is the synchrotron polarization emissivity along the LOS, and $\phi$ is Faraday depth, which is proportional to the integration of thermal electron density and magnetic ﬁelds along the LOS. From the above equation, it is clear that $F(\phi)$ and $P(\lambda^2)$ are related by Fourier transforms.
    
    The realization of Faraday tomography as a viable tool for understanding cosmic magnetism is made difficult by the limited coverage of the polarization spectrum. Firstly, measuring at negative $\lambda$ values is nonphysical, and the instrument's polarization observation coverage is fixed in practice. It is then implied that Faraday tomography is an \textit{inverse problem} that is also \textit{ill-posed}. Solving ill-posed inverse problems requires apriori information/regularization to select the best possible solution from the infinitely many solutions that satisfy the observation. 
    
    The direct inversion of the observed polarization spectra is called rotation measure (RM) synthesis \citep{Brentjens_2005}. However, to improve the reconstruction, many techniques explore the use of various apriori information. \citet{Thiebaut_2010} linearized the inversion problem with physical constraints such as the $B$ field divergence. \citet{Frick_2010} improved the reconstruction by symmetry arguments for the source along the LOS. \citet{Pratley_2020} introduced an algorithm that considers the polarization spectrum for negative $\lambda$ to be zero. \citet{Ndiritu_2021} suggested the use of Gaussian process modeling to interpolate gaps in the polarization spectrum to improve the FDF reconstruction. Meanwhile, a popular technique is $QU$-fitting \citep{Farnsworth_2011, O'Sullivan_2012, Ideguchi_2014a, Ozawa_2015, Kaczmarek_2017, Sakemi_2018, Schnitzeler_2018, Miyashita_2019}, which assumes that the FDF can be approximated by a single or a combination of simple analytic functions such as Gaussian, top-hat, or delta functions. RM \textsc{CLEAN} \citep[e.g.,][]{Heald_2009, Anderson_2016, Michilli_2018} is a matching pursuit algorithm that assumes the FDF to be a collection of point-like sources. In addition to the above, ideas of compressive sensing \citep{Donoho_2006, Candes_2006} can be used to regularize the inverse problem. \citet{Andrecut_2012} implemented a matching pursuit algorithm to fit an over-complete dictionary of functions sparsely. \citet{Li_2011} and \citet{Akiyama_2018} imposed sparsity in the Faraday depth space using regularization functions.
    
    A concept that is often used in conjunction with sparsity is wavelets. Wavelet is a wave-like function with which we can define a transform (wavelet transform) of a square integral function in terms of an orthonormal set generated by the wavelet \citep[see, e.g.][]{Daubechies_1992}. The wavelet transform is similar to the Fourier transform, where a wavelet function replaces sine and cosine functions. A vital advantage of the above transform is that the wavelet representation provides the frequency and the temporal location, allowing a scale-dependent decomposition of a function. Wavelets have also been used in Faraday tomography for the decomposition of the FDF \citep{Frick_2010, Sokoloff_2018}. An area where wavelets are extensively used is for image compression, where the signal is wavelet transformed to provide a sparse representation of the signal. For non-parametric methods that solve ill-posed problems, like Faraday tomography, the wavelet transform can significantly reduce the number of parameters that need to be estimated, ultimately improving the reconstruction of the FDF.
    
    \citet{Cooray_2021} recently introduced the \textsl{Constraining and Restoring Algorithm for Faraday Tomography} (\textsc{CRAFT}). \textsc{CRAFT} is an version of the Papoulis-Gerchberg algorithm \citep{Papoulis1975,Gerchberg1974,Cooray_2020} that imposes sparsity in Faraday depth and smoothness of the polarization angle to produce high fidelity FDF reconstructions. 
    
    This work explores the wavelet representation and its sparse nature for better reconstruction of FDF from the partially observed linear polarization spectrum. Within the ideas of compressive sensing, a sparse wavelet representation implies that only a smaller number of measurements of the linear polarization spectrum are required to contain enough information to approximate the FDF, thereby improving the reconstruction of the intrinsic FDF from partial observations. We describe the implementation of wavelet space sparsity with \textsc{CRAFT} due to the flexible nature of the algorithm in incorporating priors. 
    
    This paper is structured as follows. Section \ref{sec:technique} provides the theory and idea of the CRAFT algorithm incorporating the sparsity in wavelet representation. After that, we demonstrate the application of the presented technique on a realistic observation of an FDF in Section \ref{sec:application}. Following it, Section \ref{sec:discussion} provides some additional analysis of the results with some observations on the technique. Lastly, we conclude in Section \ref{sec:conclusion} summarizing the paper.

\section{\textsc{CRAFT} with sparsity in wavelet representation} \label{sec:technique}

    The goal of Faraday tomography is to obtain the complete linear polarization spectrum $P(\lambda^2)$ from the observed spectrum $\tilde{P}(\lambda^2)$. These quantities are related as,
    \begin{equation} \label{eq:observation}
    \tilde{P}\left(\lambda^{2}\right)= \W \left(\lambda^{2}\right) {P}\left(\lambda^{2}\right),
    \end{equation}
    where $\W \left(\lambda^{2}\right)$ is a masking operator that defines the $\lambda^2$ coverage from observation. Faraday tomography involves solving the inverse for the above equation. 
    
    \textsc{CRAFT} solves the above problem iteratively, obtaining a better solution at each successive step. The $n^{\textrm{th}}$ iteration for this algorithm is written as;
    \begin{equation} \label{eq:algorithm}
        P_{n} (\lambda^{2}) = \tilde{P} (\lambda^{2}) + \left[\I - \W (\lambda^{2}) \right] \B P_{n-1}  (\lambda^{2}),
    \end{equation}
    where $\I$ is the identity matrix, $P_{n}$ is the $n^{\textrm{th}}$ estimate of $P$, $P_0 = \tilde{P}$, and $\B$ is the regularizing operator that contains apriori knowledge of $P$. By operating $\B$ on $P_n$ at each iteration $n$, constraints are applied on the spectrum.
    
    In this work we suggest the $\B$ operator to be of the following form,
    \begin{equation} \label{eq:prior_operator}
        \B = \F \WT^{-1} \Delta_{w, \nu} \WT \Delta_{\phi,\mu} \mathsf{\beta} \F^{-1},
    \end{equation}
    where $\F$ is an operator of Fourier transform as Eq. (\ref{eq:FT}), $\mathsf{\beta}$ is a window function in Faraday depth space based on physical motivations on the largest possible value of nonzero $\phi$,  and $\WT$ is the wavelet transform defined as,
    \begin{equation} \label{eq:wavelet_transform}
        w_{F}(a, b) = \WT F(\phi) =\frac{1}{|a|} \int_{-\infty}^{\infty} F(\phi) \psi^{*}\left(\frac{\phi-b}{a}\right) \mathrm{d} \phi.
    \end{equation}
    In the above, $\psi(\phi)$ is the wavelet used for decomposition, and $w_{F}(a, b)$ the coefficients in the wavelet representation, where $a$ is the scale, and $b$ the shift parameter or roughly the location of the decomposing wavelet. 
    {The simplest wavelet that is also used in this work is the Haar wavelet. Haar wavelet can be described as,
    \begin{equation} \label{eq:Haar_def}
        \psi(\phi)=
            \left\{\begin{array}{ll}
                1 & \textrm{ if } 0 \leq \phi < \frac{1}{2}, \\
                -1 & \textrm{ if } \frac{1}{2} \leq \phi < 1, \\
                0 & \textrm{ Otherwise. }
            \end{array}\right.
    \end{equation}}
    Then the $\Delta$ operators are non-linear thresholding operators that enforce sparsity \citep{Daubechies_2004, Kayvanrad_2009}. The operator $\Delta_{\phi,\mu}$ is used to impose sparsity of the FDF in Faraday depth $\phi$ and is defined as follows,
    \begin{equation} \label{eq:delta_phi_operator}
        \Delta_{\phi,\mu}[|F(\phi)|]=\left\{
        \begin{array}{ll}
            |F(\phi)|-\mu & \textrm{ if } |F(\phi)| \ge \mu \\
            0 & \textrm{ if }|F(\phi)| < \mu
        \end{array}\right. ,
    \end{equation}
    where $\mu$ is a parameter that controls the level of sparsity in $\phi$. As a new addition to \textsc{CRAFT}, we implement a sparsity operator in the wavelet space \citep{Donoho_1994, Donoho_1995a, Donoho_1995b} defined as, 
    \begin{equation} \label{eq:delta_w_operator}
        \Delta_{w, \nu}[w_{F}(a, b)]=
            \left\{\begin{array}{ll}
                w_{F}(a, b)+\nu & \textrm{ if } w_{F}(a, b) \leq-\nu \\
                0 & \textrm{ if } |w_{F}(a, b)|<\nu \\
                w_{F}(a, b)-\nu & \textrm{ if } w_{F}(a, b) \geq \nu
            \end{array}\right. ,
    \end{equation}
    with $\nu$ being the parameter that controls the sparsity of $w_F$. By setting small wavelet coefficients to zero, we are essentially removing wavelet components that have minimal effect on the overall shape of the FDF. This process is commonly referred to as wavelet shrinkage (WS). 
    Thus, increasing the sparsity of wavelet coefficients $w_F$ by controlling the threshold parameter $\nu$ can reduce the number of features captured by the chosen wavelet. Because the Fourier transform can be considered within the larger class of wavelet transforms, it may appear redundant to compute first the Fourier and then the wavelet transform. However, computing the (inverse) Fourier transform from the observed linear polarization spectrum to obtain the FDF provides a vital purpose; it allows the application of wavelet shrinkage on the physically meaningful representation of data, i.e., FDF.
    
    The above process in Eq. (\ref{eq:algorithm}) is iterated until a convergence criterion is met to produce the reconstructed linear polarization spectrum, from which the reconstructed FDF is obtained. An example of a convergence criterion is the relative residual, i.e.,  $||P_{n}-P_{n-1}||_{2}/||P_{n-1}||_{2} < \epsilon$, where  $||P_{n-1}||_{2}$ is the $\ell_2$ norm of $P_{n-1}$ and $\epsilon$ is some small positive number. The appropriate values for parameters $\mu$ and $\nu$ can be determined through a grid search with an appropriate condition. CRAFT is a version of projected gradient descent, and thus we expect each estimation to move towards the groundtruth \citep{Combettes_2009}. Please see \citet{Cooray_2021} for a detailed discussion on the algorithm.
    
\begin{figure}
        \centering
        \includegraphics[width=0.7\linewidth]{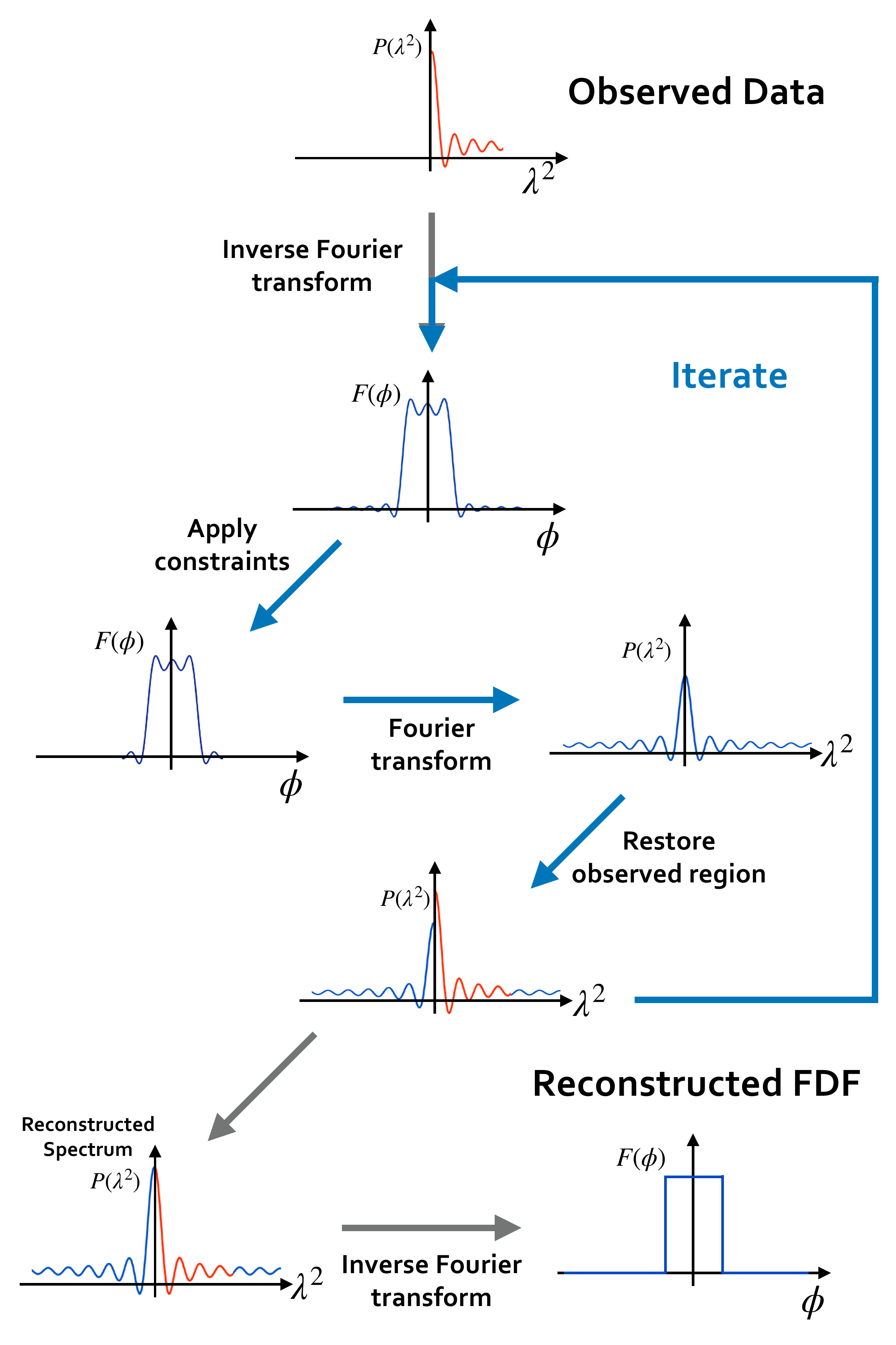}
        \caption{A diagram explaining the general procedure of the \textsc{CRAFT} algorithm. The process begins with the inverse Fourier transform of the linear polarization spectrum to obtain an FDF. The constraints (Eq. \ref{eq:prior_operator}) are applied on this FDF as sparsity in $\phi$ as Eq. (\ref{eq:delta_phi_operator}) and wavelet shrinkage in FDF amplitude and polarization angle (Eq. \ref{eq:delta_w_operator}). After the constraints, the FDF is Fourier transformed back to obtain as estimate a linear polarization spectrum. Due to the constraints, the obtained polarization spectrum is different from the original, and parts of the unobserved regions are reconstructed. The observed regions of the linear polarization spectrum are restored to obtain the first estimate of the reconstructed linear polarization spectrum. The above procedure is repeated until the convergence criterion is met to obtain the reconstructed linear polarization spectrum. After that, to obtain the final reconstructed FDF, the reconstructed linear polarization spectrum is inverse Fourier transformed\protect\footnotemark[1].} 
        \label{fig:algorithm}
    \end{figure}

    Figure \ref{fig:algorithm} is provided for easy understanding of the general \textsc{CRAFT} algorithm. Hereafter, the new method presented in this paper will be referred to as \textsc{CRAFT} + WS to distinguish from the original \textsc{CRAFT} algorithm presented in \citet{Cooray_2021}. We note that in practice, we employ the discrete versions of Fourier and wavelet transforms. The continuous wavelet transform samples the scale and shift parameters continuously, creating redundancies. The discrete wavelet transform provides an extremely efficient and practical way to transform real-world signals like FDFs. Our implementation that follows uses the wavelet transforms of \textsc{PyWavelets} \citep{Lee_2019}.
    
    \footnotetext[1]{Fourier and inverse Fourier transforms are used to refer to the transform defined in Eq. (\ref{eq:FT})}

\section{Application: Reconstructing a Realistic Observation} \label{sec:application}

    We demonstrate the proposed method for a synthetic FDF in \citet{Ideguchi_2014b} of a realistic simulation of the Milky Way \citep{Akahori_2013}.  The synthetic model FDF is a mixture of both Faraday-thin ($\lambda^2 \Delta \phi \ll 1$) and thick ($\lambda^2 \Delta \phi \gg 1$) components, where $\Delta \phi$ is the extent of the source in $\phi$. The complicated FDF is ideal for testing the non-parametric techniques' ability to capture multi-scale information in $\phi$. Additionally, reconstructions for same model FDF are also shown in \citet{Akiyama_2018} and \citet{Cooray_2021}. For further descriptions of the model, see Appendix \ref{sec:appendix-1}

    The model FDF has a $\phi$ range of -1000 to 1000 [rad m$^{-2}$] with a resolution of 0.1 [rad m$^{-2}$]. The FDF is numerically Fourier transformed to obtain the linear polarization spectrum. A part of the spectrum is adopted as the simulated observation, considering three cases of frequency coverage. The first is the observation frequency ranges of Australian Square Kilometre Array Pathfinder \citep[ASKAP;][]{ASKAP}, which corresponds to the range of 700 [MHz] to 1800 [MHz]. Secondly, we consider the case of Square Kilometre Array (SKA) Phase 1 mid-frequency bands (Bands 1 \& 2) that correspond to the frequency range of 350 [MHz] to 1760 [MHz]. Lastly, we consider the widest continuous coverage, adding SKA LOW to the second scenario. SKA LOW + MID corresponds to the frequency range of 50 [MHz] to 1760 [MHz]. The choice of these three frequency coverage cases demonstrates the potential of the introduced technique to upcoming large (all-sky) surveys used for Faraday tomography. The three frequency ranges; 700 [MHz] - 1800 [MHz], 350 [MHz] - 1760 [MHz], 50 [MHz] - 1760 [MHz], will be called Case 1, Case 2, and Case 3, respectively from here onward. The number of samples in the observed linear polarization spectrum $\tilde{P}(\lambda^2)$ are 99, 499, and 2524 for cases 1, 2, and 3, respectively. After limiting the frequency coverage of the data, frequency-independent random Gaussian noise with the zero mean and the standard deviation of 0.1 [mJy] is added to each $\lambda^2$ channel of Stokes $Q$ and $U$. We consider every sampling of $P(\lambda^2)$ infinitesimally narrow and spaced equally in $\lambda^2$ space. 

    The algorithm introduced in Section \ref{sec:technique} is implemented for the above setup as follows. The most crucial detail is in the $\B$ operator at each iteration step. At iteration $i$, we perform the inverse Fourier transform to obtain an estimate for the FDF. Then, a non-linear thresholding operator $\Delta_{\phi,\mu}$ in Eq. (\ref{eq:delta_phi_operator}) is applied to the FDF to ensure sparsity in $\phi$, where amplitudes values smaller than $\mu$ are set to zero. The small value $\mu$ is subtracted from the rest of the FDF amplitude. After that, we implement wavelet shrinkage on the FDF. Wavelet shrinkage on FDF is applied separately on the amplitude and the phase as they tend to have different intrinsic shape characteristics. The varying amplitude and phase features suggest the use of separate wavelet families when transforming them. 
    
    Choosing the decomposing wavelet is no trivial task. In this work, we propose to use the most straightforward Haar wavelet as the wavelet to regularize the polarization angle reconstruction. Haar wavelet introduced above has a step like feature that can scale to fit multi-scale variations in the signal. Shrinkage of the Haar wavelet coefficients will cause the resultant function to have fewer "steps". 
    Such a polarization angle shape is typical even in parametric model fitting (\textit{QU}-fitting), where simple sources with each having a constant polarization angle are added together to fit the $QU$ data. Additionally, the use of constant polarization angle for a Faraday source supports the symmetry arguments described in \citet{Frick_2010}, where sources are assumed to be symmetric along the LOS.
    
    In the case of amplitude, we use the coiflet wavelet family. Coiflet wavelet is a nearly symmetric wavelet that has been commonly used for decomposing spectra like signals that contain peaks as well as a smooth component \citep[e.g.,][]{Donoho_1995, Srivastava_2016}. 
    In the coiflet wavelet family, we need to determine the number of vanishing moments or the approximation order of the decomposing wavelet. The shape of the decomposing wavelet changes depending on the vanishing moments $N$ even within the same family. The wavelet shape is described by a set of vales. For example, the simplest Coiflet wavelet ($N=1$) is described by the coefficients -0.07273, 0.33790, 0.85257, 0.38486, -0.07273, -0.01566. A common consensus is that higher vanishing moments relate to a higher compression capability, but the ideal approximation order is not trivial and depends on the application. We skip the technical details in this paper and ask the reader to follow many available reviews on wavelets \citep[e.g.,][]{Dremin_2001}. Here, we consider the sparsity in representing the FDF as a criterion of the best wavelet within its family. The best wavelet to decompose the FDF will be used in the "Apply constraints" step in the Figure \ref{fig:algorithm}. To determine the best wavelet for the application, we first inverse Fourier transform the observed linear polarization spectrum to obtain the observed FDF. The obtained FDF is wavelet transformed with each coiflet wavelet degree that is available in \textsc{PyWavelets}. Then the $\ell_2$ norms of each wavelet representation are calculated. The wavelet with the smallest $\ell_2$ norm will give the most sparse representation of the resultant FDF. In practice, the approximation order $N$ should depend on the complexity of the intrinsic FDF, observed frequency coverage, and the resolution in Faraday depth space.
    
\begin{figure*}[ht]
        \centering
        \includegraphics[width=\linewidth]{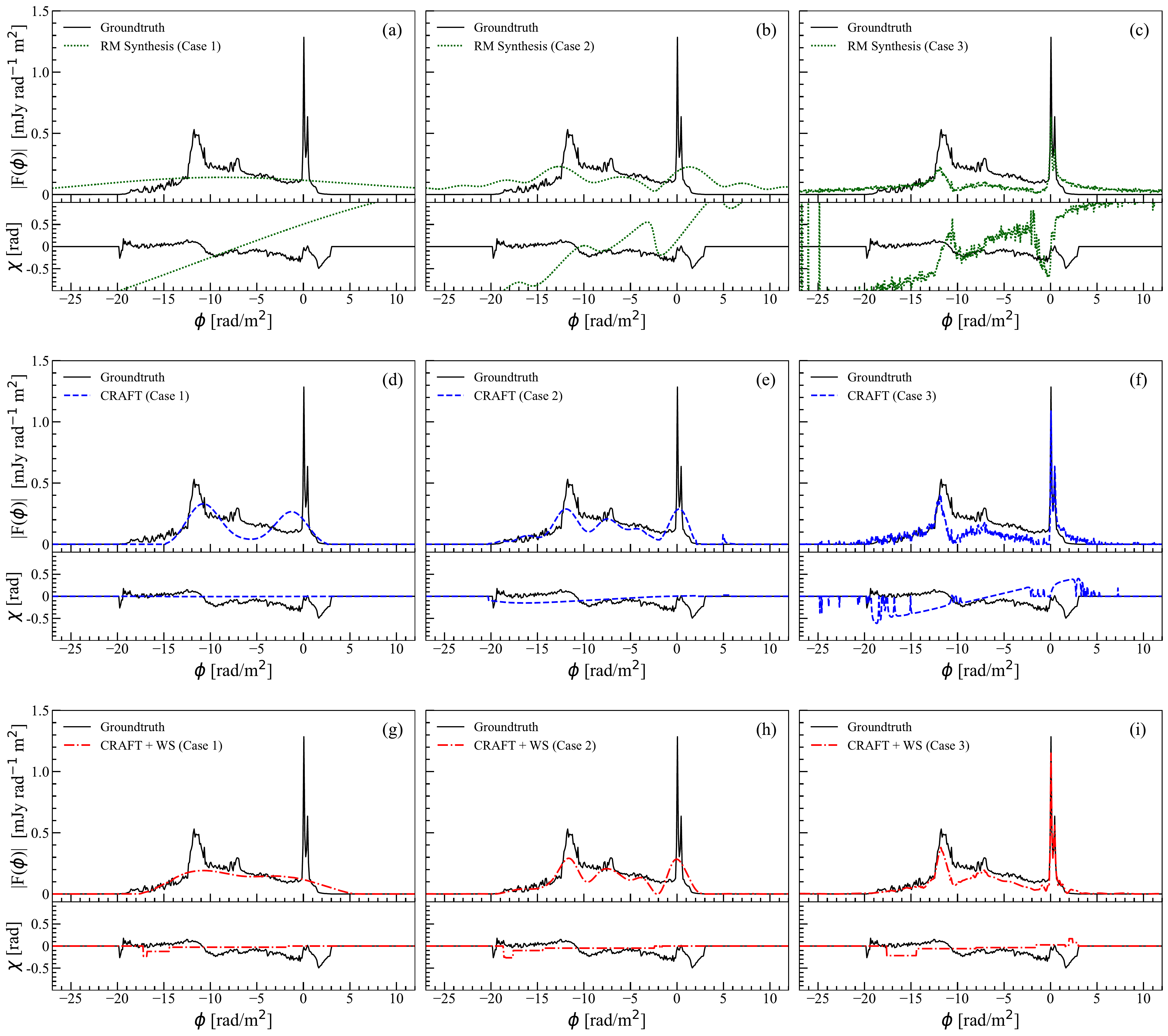}
        \caption{A comparison of reconstructions of a realistic galaxy FDF for observations with 3 frequency coverage cases. The three columns corresponds to 700 [MHz] - 1800 [MHz], 350 [MHz] - 1760 [MHz], and 50 [MHz] - 1760 [MHz], frequency ranges, respectively. Black solid line is the original model FDF (noiseless) and the three rows from the top correspond to the reconstruction with RM Synthesis (green dotted), \textsc{CRAFT} (blue dashed), and the new technique proposed in this work, \textsc{CRAFT} + WS (red dash dotted). In each panel, the upper part shows the amplitude and the bottom part shows the polarization angle.}
        \label{fig:all_reconstruction}
    \end{figure*}

    
    We demonstrate the results of the standard RM synthesis \citep{Brentjens_2005}, CRAFT \citep{Cooray_2021} and \textsc{CRAFT} + WS (this work) in Figure \ref{fig:all_reconstruction} for the three cases of frequency coverage. The three rows correspond to the result using three techniques used for reconstruction. From the top, they are RM Synthesis, CRAFT, and CRAFT + WS. The three columns corresponds to the three cases of frequency coverage, i.e., 700 [MHz] - 1800 [MHz], 350 [MHz] - 1760 [MHz], and 50 [MHz] - 1760 [MHz]. RM synthesis is the simplest inverse Fourier transform (Eq. \ref{eq:FT}) of the observed linear polarization spectrum with the unobserved regions set to zero. All the reconstructions had the Faraday depth bin width of 0.1 [rad m$^{-2}$] (same as the original model). Visually, it is clear that the reconstruction improves as we go down the three techniques, becoming closer to the original model FDF. It is visible that as the frequency coverage increases, we start seeing the reproduction of finer structures.
    
\begin{figure*}
        \centering
        \includegraphics[width=\linewidth]{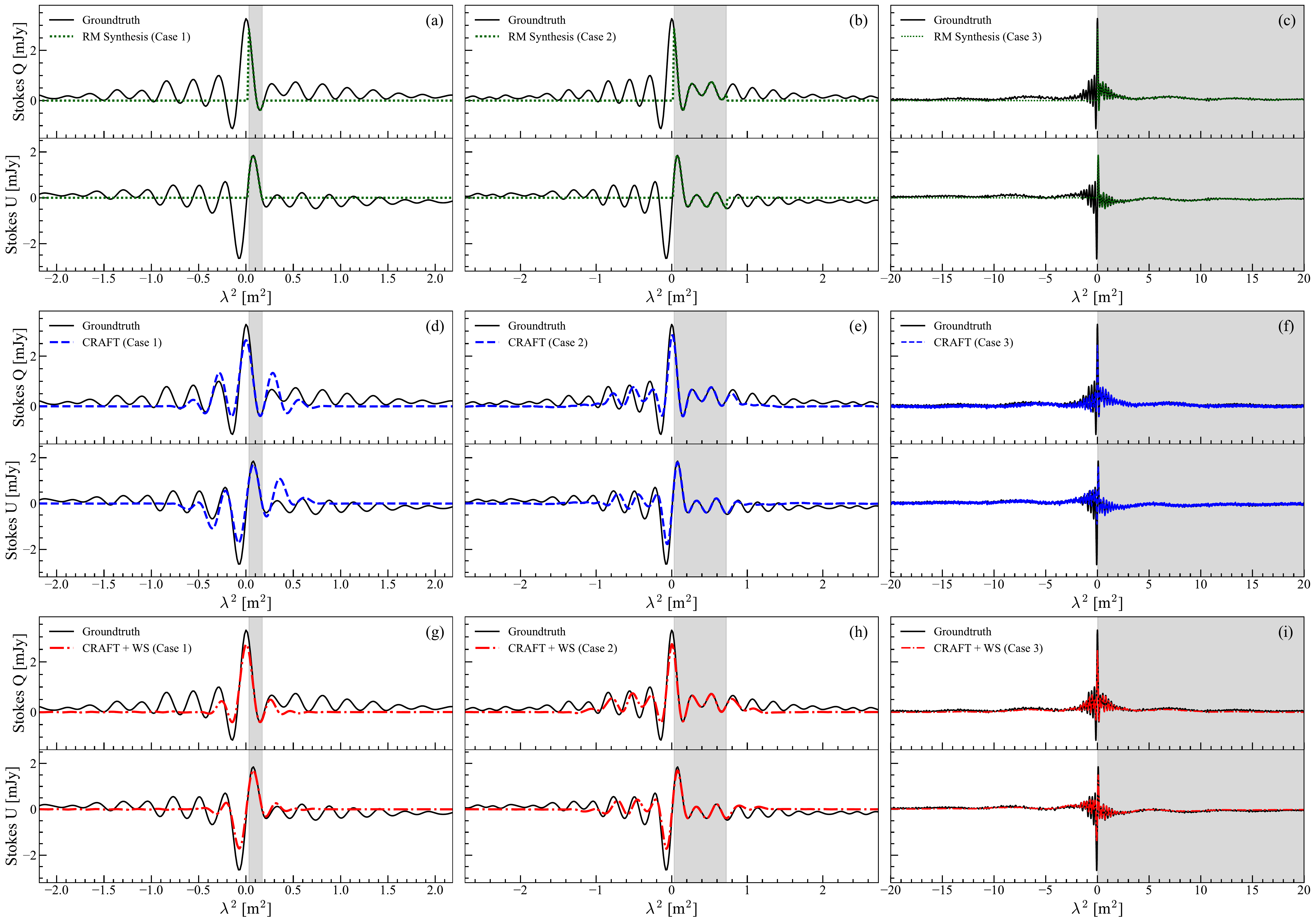}
        \caption{Corresponding linear polarization spectrum reconstructions shown in Figure \ref{fig:all_reconstruction}. The gray shaded regions corresponds to the observed frequency coverage for each coverage cases. Coverage cases 1 to 3 corresponds to ($\lambda^2_{\mathrm{obs, min}}$,  $\lambda^2_{\mathrm{obs, max}}$) = (0.0277 [m$^2$], 0.1834 [m$^2$]), (0.0290 [m$^2$], 0.7336 [m$^2$]), and (0.0290 [m$^2$], 35.9502 [m$^2$]), respectively.}
        \label{fig:all_reconstruction_P}
    \end{figure*}

    We note that both CRAFT methods that employ sparsity constraints in Faraday depth successfully correct the spreading effect seen in the RM synthesis FDF out from the source region in $\phi$. As a consequence of using wavelet shrinkage of the FDF amplitude, we do not see the noise-like features in CRAFT + WS result, which are present in both RM Synthesis and CRAFT. {Reconstruction improvements is seen} in the polarization angle, where there is better large-scale agreement going from RM Synthesis to CRAFT and then to CRAFT + WS. {Though there is a better agreement for $\chi$ with CRAFT methods, we note that the two methods cannot fully reproduce polarization angle in Faraday tomography due to the lack of negative $\lambda^2$ coverage in observations. $P(\lambda^2)$ in the negative $\lambda^2$ regime is crucial in constraining the phase (polarization angle) of the FDF as clearly demonstrated in Figure \ref{fig:all_reconstruction_P}. Reconstruction errors in FDF amplitude (especially between $\phi \approx$ -12 [rad/m$^2$] to $\phi \approx$ 0 [rad/m$^2$]) can also be attributed to the same reason as above, where depolarization has not been completely solved. Reconstruction residuals for each frequency coverage is shown in Appendix \ref{sec:appendix-2}.}
    
    Figure \ref{fig:all_reconstruction_P} shows the corresponding linear polarization spectra for the reconstruction shown in Figure \ref{fig:all_reconstruction}. The gray regions demonstrate which parts of the spectrum are observed and which parts are extrapolated by reconstruction. RM synthesis fills the unavailable regions of $P(\lambda^2)$ with zeros, whereas CRAFT and CRAFT+WS have managed to reproduce certain parts of the spectrum to a limited degree. By constraining the polarization angle of the FDF, values at negative $\lambda^2$ can be constrained. However, in this way, only the spectrum between ($-\lambda^2_{\mathrm{obs, max}}$,  $-\lambda^2_{\mathrm{obs, min}}$) can be constrained. Since $\lambda^2_{\mathrm{obs, min}} \approx 0$ for most Faraday tomography observations, CRAFT and CRAFT+WS can produce non zero values for $P(\lambda^2)$ roughly between $-\lambda^2_{\mathrm{obs, max}}$ and $\lambda^2_{\mathrm{obs, max}}$. Despite the possibility, we stress again that knowing the complete polarization angle of FDF apriori is impossible.

    The parameters for reconstruction with CRAFT and CRAFT + WS were determined with considerations of; (1) the sparsest FDF solution in each representation should be adopted, (2) the difference between fitted points and the observed points in $\lambda^2$-space should be minimal, (3) the integrated intensity in $\phi$-space should be conserved in reconstruction. Since the algorithm naturally leads to the least square solution, the parameters were selected by a grid search based on the highest thresholding values with $||F_{\mathrm{recon}}||_2 / ||\tilde{F}||_2 \approx 1$, where $F_{\mathrm{recon}}$ is the reconstructed FDF, and $\tilde{F}$ is the RM Synthesis FDF.
    
    The selected parameters, with the iterations till convergence and the reconstruction error is summarized in Table \ref{table:parameters}. The parameters $\nu_{\mathrm{amp}}$, and $\nu_{\mathrm{ang}}$ are the two parameters that control the sparsity of the FDF amplitude and the polarization angle in their respective wavelet representation. The metric used to quantify the reconstruction error is the the normalized root mean squared error \citep[NRMSE;][]{Fienup_1997}. We define the NRMSE for reconstructed FDF with respect to the original as,
    \begin{equation}
        \mathrm{NRMSE}(\hat{F}, F)=\sqrt{\frac{\sum_{i}\left(\hat{F}_{i}-F_{i}\right)^{2}}{\sum_{i}\left(F_{i}\right)^{2}}},
        \label{eq:nrmse}
    \end{equation}
    where $\hat{F}$ is the reconstructed complex FDF and the $F$ is the model FDF. We see that generally we see a drastic improvement in the reconstruction for both CRAFT methods when compared to RM Synthesis. In particular, CRAFT + WS (this work) demonstrate additional improvements over the CRAFT method.
    
    \def\arraystretch{1.1}%
    {\setlength{\tabcolsep}{1.5pt}
    \begin{table}
    \centering
        \begin{tabulary}{\linewidth}{l c l c}
        Method & Iterations & Parameters       & NRMSE  \\ \hline \hline
        \multicolumn{2}{l}{\textbf{Case 1} (700 MHz - 1800 MHz)}           &                  &        \\
        RM Synthesis          & -          & -                & 0.9669 \\
        CRAFT                 & 1000       & 0.01             & 0.6264 \\
        CRAFT + WS            & 678        & 0.01, 0.5, 0.002 & 0.6038 \\
                              &            &                  &        \\
        \multicolumn{2}{l}{\textbf{Case 2} (350 MHz - 1760 MHz)}            &                  &        \\
        RM Synthesis          & -          & -                & 0.9609 \\
        CRAFT                 & 527        & 0.005            & 0.5661 \\
        CRAFT + WS            & 178        & 0.01, 0.5, 0.005 & 0.5182 \\
                              &            &                  &        \\
        \multicolumn{2}{l}{\textbf{Case 3} (50 MHz - 1760 MHz)}            &                  &        \\
        RM Synthesis          & -          & -                & 0.8189 \\
        CRAFT                 & 370        & 0.02             & 0.6595 \\
        CRAFT + WS            & 53         & 0.02, 0.8, 0.02  & 0.4054 \\
        \hline
        \end{tabulary}
        \caption{The iteration till convergence, grid search selected parameters, and the NRMSE for the various cases of FDF reconstruction. RM Synthesis does not require any parameters for reconstruction, while CRAFT and CRAFT + WS (this work) requires some additional parameters. The parameter shown for CRAFT is $\mu$ and for CRAFT + WS, the parameters shown are $\mu$, $\nu_{\mathrm{amp}}$, and $\nu_{\mathrm{ang}}$, respectively. NRMSE decreases with increasing frequency range for all the techniques. CRAFT performs better than RM Synthesis, and CRAFT + WS performs the best in terms of NRMSE (lower the better).
        }
        \label{table:parameters}
    \end{table}}
    
    There are two points to be noted. First, the iterations required until convergence are lower for CRAFT + WS than CRAFT. The faster convergence can be attributed to the degree of freedom for estimation being limited to fewer nonzero coefficients in wavelet space compared to all the points in Faraday depth space for CRAFT. Second, while RM synthesis does not require any additional parameters for reconstruction, CRAFT methods require parameters to be decided. The new CRAFT + WS technique requires two additional parameters for controlling the FDF amplitude and phase wavelet sparsity in addition to the $\mu$ parameter required by the original CRAFT.

\section{Discussion} \label{sec:discussion}

    To quantitatively assess the reconstruction performance at each resolution in $\phi$, we perform a multi-scale error analysis using the same error metric used above, NRMSE. The error analysis is performed by computing the smoothed version of the reconstructed and the model FDFs and calculating the error metric. We can further understand the reconstruction behavior and the performance at various resolutions in $\phi$. The smoothing is done with a Gaussian kernel with full-width-at-half-maximum (FWHM) corresponding to 0.1 to 10 times the FWHM of the rotation measure spread function (RMSF). According to Equation (61) of \citet{Brentjens_2005}, the FWHM of the RMSF depends on the available frequency coverage as {approximately} $2\sqrt{3}/(\lambda^2_{\textrm{obs, max}} - \lambda^2_{\textrm{obs, min}})$, where $\lambda^2_{\textrm{obs, min}}$ and $\lambda^2_{\textrm{obs, max}}$ are the minimum and the maximum $\lambda^2$ values in the observation coverage. The above equation assumes top hat weight function (one for observed and zero elsewhere), which is used in this paper. For the the top hat weight function, RMSF FWHM becomes the above form. The RMSF FWHM values for the three cases are 22.25 [rad m$^{-2}$], 4.91 [rad m$^{-2}$], and 0.10 [rad m$^{-2}$] for 700 [MHz] - 1800 [MHz], 350 [MHz] - 1760 [MHz], and 50 [MHz] - 1760 [MHz], respectively. 
    
\begin{figure}
        \centering
        \includegraphics[width=0.45\textwidth]{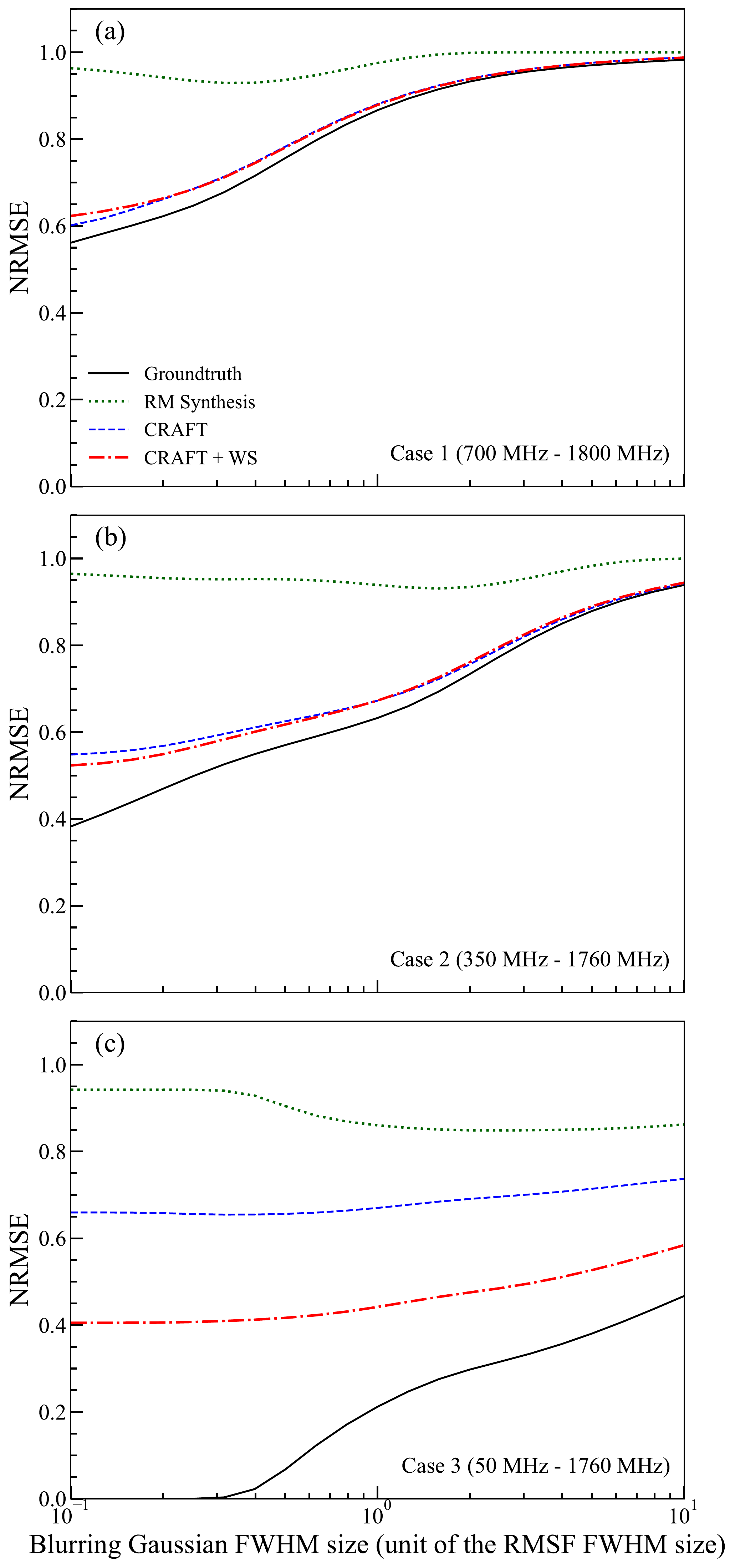}
        \caption{Results of the multi-scale error analysis. The FDFs are smoothed with a Gaussian kernel of size corresponding to a multiple of the RMSF FWHM. Then NRMSE is calculated with respect to the smoothed FDF's and the original model. The three panels from the top correspond to the three cases of frequency coverage. The solid black line corresponds to the original model FDF, the green dotted line is for RM Synthesis, the blue dashed line for CRAFT, and the red dash-dotted line for CRAFT + WS (this work). Lower is better for NRMSE.}
        \label{fig:multiscale_analysis}
    \end{figure}

    The result of the multi-scale error analysis is shown in Figure \ref{fig:multiscale_analysis}. We expect error values to be bounded by the RM Synthesis line (green dotted) at the top and the original model line (black solid) at the bottom for any appropriate Faraday tomography technique. Notice that even the original model has nonzero NRMSE. The discrepancy is that when a Gaussian smooths the original model, the FDF loses information of the scales smaller than the smoothing scale. The best Faraday tomography techniques should be closest to the original model line at every $\phi$ scale in the multi-scale error analysis plot. It is seen that both CRAFT methods lie closer to the original model line in comparison to RM synthesis, confirming the capability for accurate reconstruction. Additionally, we observe that CRAFT + WS generally offer noticeable improvements over the original CRAFT method.
 
    The multi-scale error analysis also assesses the ability for super-resolution. Super-resolution in this context is to reconstruct on $\phi$ scales smaller than the FWHM of the RMSF. As seen in Figure \ref{fig:multiscale_analysis}, we do not observe a significant improvement in the FDF reconstruction for scales smaller than the RMSF FWHM for RM Synthesis. However, we see a continued decrease in the error metric for smaller scales for the CRAFT methods, suggesting that CRAFT methods can achieve super-resolution in Faraday depth. CRAFT and CRAFT + WS achieve super-resolution by imposing constraints on the polarization angle, thereby effectively reconstructing the linear polarization spectrum's negative $\lambda^2$ side. For more details on this discussion, please refer to \citet{Cooray_2021}. The above capability of the super-resolution suggests that even if one may observe a source as Faraday simple, CRAFT, and CRAFT + WS can uncover hidden complexities in the FDF.
    
    We identify some critical advantages of incorporating wavelet shrinkage to Faraday tomography. Wavelets provide a more flexible way to regularize the polarization angle reconstruction. Considering the Fourier transform, constraining the polarization angle of the FDF is difficult when there are no observations in the negative $\lambda^2$ side of $\tilde{P}\left(\lambda^2\right)$. The original CRAFT attempted to smooth out the reconstructed polarization angle for scales smaller than the FWHM of the RMSF. 
    However, in that case, the smoothing scale can be too small to impose any meaningful apriori information depending on the frequency coverage as seen for Case 3. 
    In CRAFT + WS, shrinking the Haar wavelet coefficients of the polarization angle allows for a step-wise constant polarization angle that is not dependent on the RMSF FWHM scale. 
    The added benefit is that we can control the number of steps by controlling the $\nu_{\mathrm{ang}}$ parameter, providing us a versatile and flexible way of imposing constraints on the polarization angle.
    
    The other key benefit of utilizing wavelet shrinkage is on the FDF amplitude. Wavelet representation of a signal is often more sparse given the choice of an appropriate decomposing wavelet. The sparsity in the wavelet space allows us to represent a complex signal with relatively fewer coefficients. In Faraday tomography, where the information of the FDF is limited, the sparsity helps with the ill-posedness of the inversion problem. \citet{Andrecut_2012} uses an over-complete dictionary of functions and sparsity to overcome the ill-posedness. Compared to the older version of CRAFT, wavelets speed up the convergence rate because of the smaller degree of freedom in the sparse wavelet representation. In other words, sparse representation means that fewer nonzero values need to be determined for the solution resulting in faster convergence. However, the above benefit comes at the expense of computing forward and backward wavelet transforms in addition to the forward and backward Fourier transforms at each iteration present in CRAFT. Despite the added computational complexity, CRAFT + WS often finished in about half the time as the original CRAFT with the same experimental setup because of the significant reduction in the number of iterations till convergence (see Table \ref{table:parameters}). Wavelet shrinkage also benefits in denoising the FDF, as mentioned above. 
    
    
    An essential consideration for WS is the choice of decomposing wavelet. For demonstration purposes, we used the coiflet and Haar wavelets. Determining a suitable wavelet is not easy and often done through trial and error. Thus, it is worth exploring other well-known wavelets or designing a custom wavelet for Faraday tomography. However, deriving a universal characteristic wavelet for Faraday tomography is challenging as even for similar physical properties, the FDF shape is highly dependent on the configuration of turbulence \citep{Ideguchi_2014b}. One can also use Gaussian functions as a wavelet \citep{Tsukakoshi_2012, Gossler_2021}, to imitate fitting multiple Gaussian functions as done in $QU$-fitting with this algorithm.

    FDFs often are seen as with one or two Faraday screens \citep{Anderson_2015} and the rest can be explained by three components \citep{O'Sullivan_2017}. Accurate classifications of whether a detected FDF is Faraday simple or complex was recently achieved through machine learning techniques in \citet{Brown_2018,Alger_2021}. However, the reason that FDFs are mostly Faraday simple is likely that they could not be resolved in $\phi$ due to the insufficient sampling in frequency, despite them being intrinsically Faraday complex \citep{Alger_2021}. Complex FDFs are also the most interesting, as they uncover detailed information about the intervening magnetoionic structures such as the Galactic interstellar medium \citep{Anderson_2015}. 
    For resolving these complex FDFs, we cannot depend on the improvement of the observational instruments because we can never have the full linear polarization spectrum. 
    {It is impossible to observe electromagnetic waves of imaginary frequencies (corresponds to negative squared wavelength) though they are required mathematically to obtain the complete FDF. The above point is highlighted by reconstructing Case 3 frequency coverage (SKA LOW + MID). Case 3 corresponds to an RMSF FWHM of 0.1 [rad m$^{-2}$], which is the same as the resolution of the original model. The lack of perfect reconstruction indicates that even if the smallest $\phi$-scale information is available, not knowing the negative $\lambda^2$ of $P\left(\lambda^2\right)$ will hamper Faraday tomography.} Therefore, we must explore novel reconstruction techniques to ensure the success with the existing and upcoming radio telescopes such as MeerKAT \citep{MeerKAT}, Low Frequency Array \citep[LOFAR;][]{LOFAR}, the Murchison Widefield Array \citep[MWA;][]{MWA}, ASKAP \citep{ASKAP}, and the Karl G. Jansky Very Large Array \citep[JVLA;][]{VLASS_2020}.

\section{Conclusion} \label{sec:conclusion}
    
    This paper introduced a novel model-independent reconstruction technique for Faraday tomography with the use of wavelets and sparsity. The new technique, named CRAFT + WS, was tested along with RM Synthesis and the original CRAFT on a simulated FDF \citep{Ideguchi_2014b} of a sophisticated Milky Way model \citep{Akahori_2013}. The test included simulation for frequency coverage that represents current and upcoming radio telescopes such as ASKAP and the SKA. The results suggest that CRAFT + WS can outperform RM Synthesis and CRAFT in producing the closest to the original model FDF. A multi-scale error analysis was performed on the reconstructed FDFs, which confirmed that wavelet sparsity could be a viable technique for producing the best results in Faraday tomography. 
    
    We summarize the key ideas employed to improve the FDF reconstruction from the observed partial linear polarization spectrum. They are; 
    \begin{itemize}
        \item \textbf{Sparsity of the FDF in Faraday depth} - complex polarized intensity accumulates in Faraday depth like a random walk, meaning that nonzero intensity should be observed in a confined region of Faraday depth 
        \item \textbf{Polarization angle regularization} - imposing that parts or all of the nonzero FDF have constant polarization allows us to reconstruct the negative $\lambda^2$ regions of the linear polarization spectrum, improving the resolution in Faraday depth
        \item \textbf{Regularization of the FDF amplitude} - limiting the degree of freedom on the shape to improve the ill-posedness of the Faraday tomography problem
    \end{itemize}
    The above apriori information was implemented within a version of the projected gradient descent algorithm (CRAFT) as thresholding operators of the Faraday depth space and wavelet representation FDF amplitude and polarization angle.

    We also argue that CRAFT+WS demonstrates clear advantages in computational efficiency and performance under significant noise. However, we have skipped the demonstrations explicit testing of these features in this paper. Comprehensive testing of the reconstruction methods will be provided in the companion paper of Ideguchi et al. in preparation. CRAFT and CRAFT + WS codes will be made publicly available\footnotemark[3].

\section{Funding}
    SC is supported by the Japan Society for the Promotion of Science (JSPS) under Grant No. 21J23611. This work was supported in part by JSPS Grant-in-Aid for Scientific Research (TTT: 19H05076, 21H01128, TA: 21H01135). TTT is supported in part by the Sumitomo Foundation Fiscal 2018 Grant for Basic Science Research Projects (180923), and the Collaboration Funding of the Institute of Statistical Mathematics ''New Development of the Studies on Galaxy Evolution with a Method of Data Science''. KT is partially supported by JSPS KAKENHI Grant Numbers 20H00180, 21H01130, and 21H04467, Bilateral Joint Research Projects of JSPS, and the ISM Cooperative Research Program (2021-ISMCRP-2017).

\section{Acknowledgment}
We thank the anonymous referee for their careful reading of the manuscript to provide suggestions that significantly improved this paper.




\section{References}
\bibliographystyle{bibstyle}
\renewcommand{\bibsection}{}
\bibliography{references}



\appendix
\section{Simulated Galaxy Model} \label{sec:appendix-1}
    
\begin{figure}
        \centering
        \includegraphics[width=0.45\textwidth]{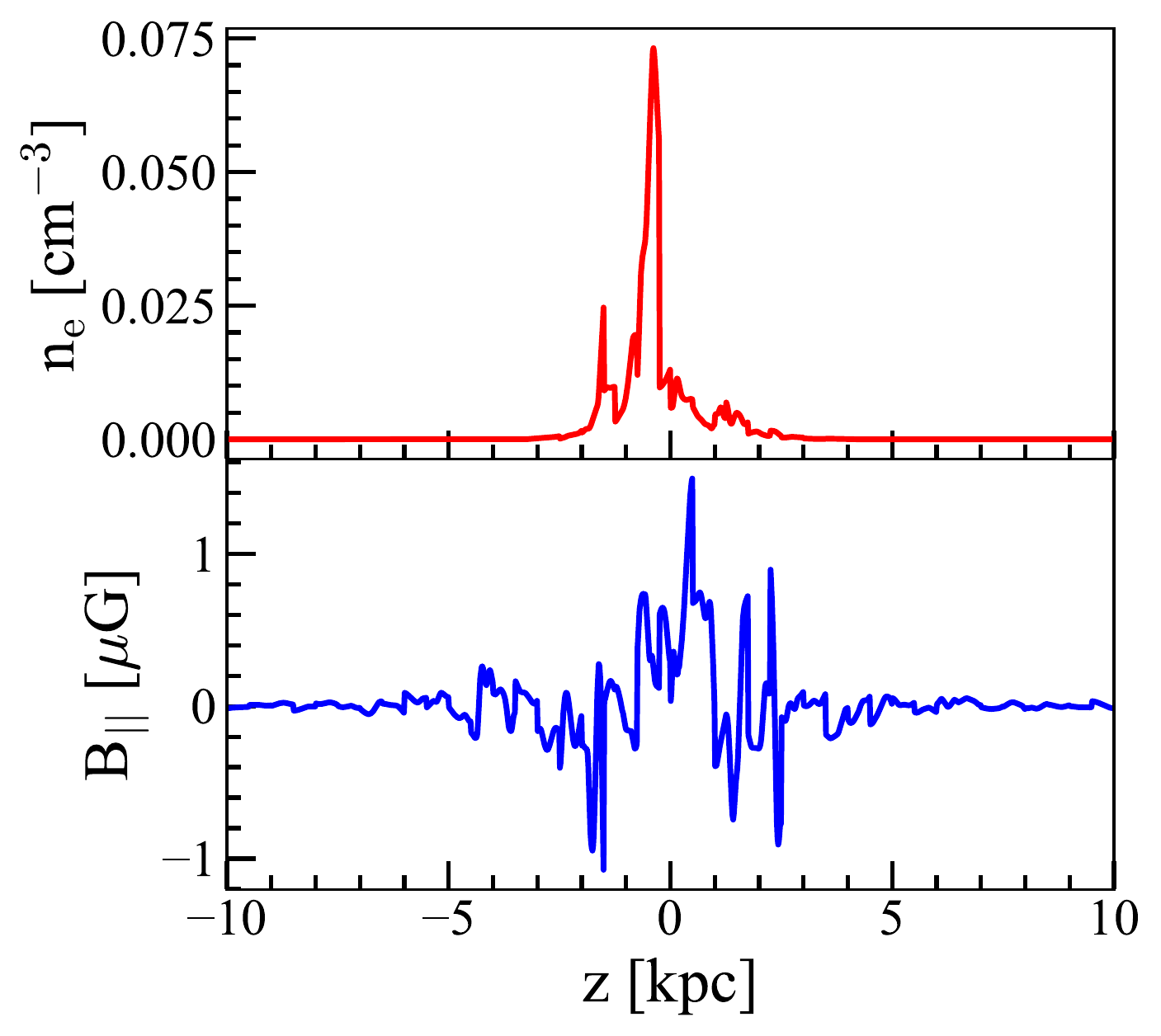}
        \caption{LOS (z-axis) distributions of thermal electron density $n_e$ of the galaxy model (red solid line) and component of magnetic field parallel to the LOS (blue solid line). }
        \label{fig:distributions}
    \end{figure}

    The FDF model used in this paper is derived based on \citet{Akahori_2013}, which incorporates both the regular global component and the turbulent random component for a spiral galaxy. The global   magnetic field includes an axisymmetric spiral field together with a halo toroidal field \citep{Sun_2008}. The cosmic-ray electric densities are modeled by an exponential model \citep{Sun_2008};
    \begin{equation}
        C(z)=0.0001 [\mathrm{cm}^{-3}] \ \mathrm{exp}(-|z|/1 \mathrm{[kpc]}),
    \end{equation}
    where $z$ is the z direction perpendicular to the galactic plane and also considered the LOS. The thermal electron densities are modeled as \citet{Cordes_2002};
    \begin{equation}
        n_e(z)=0.02 [\mathrm{cm}^{-3}] \ \mathrm{sech}^2(z/1 \mathrm{[kpc]}).
    \end{equation}
    The turbulence is modeled by piling up simulation boxes of 500 [pc] with 512 grids a side. The considered turbulence is of driving scale 250 [pc] with a root mean square flow speed of 30 [Km s$^{-1}$]. Figure \ref{fig:distributions} show the thermal electron and parallel component of the magnetic field along the LOS for the model used in this paper. It is clear that large fluctuations occur in both thermal electron densities and magnetic fields due to the presence of turbulence. See \citet{Ideguchi_2014b} for further details.

\section{Reconstruction residuals} \label{sec:appendix-2}

\begin{figure}
        \centering
        \includegraphics[width=0.45\textwidth]{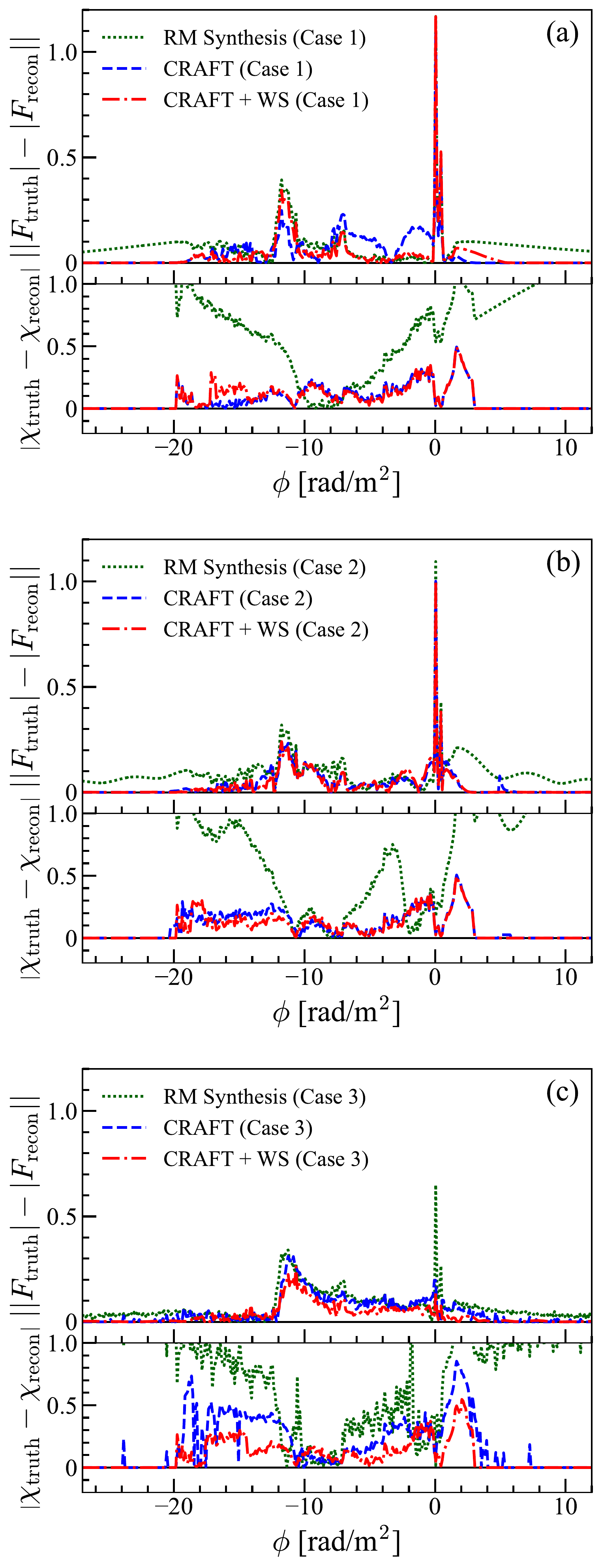}
        \caption{Absolute reconstruction residuals for the FDF amplitude and polarization angle. Zero implies the best reconstruction. Panel (a) is for Case 1 (700 [MHz] - 1800 [MHz]), panel (b) for Case 2 (350 [MHz] - 1760 [MHz]), and panel (c) for Case 3 (350 [MHz] - 1760 [MHz]). In each panel, the upper half shows the residual in amplitude and lower half shows the polarization angle residual. Green dotted line is the absolute reconstruction residual for RM synthesis, blue dashed line is for CRAFT and red dash-dotted line is for CRAFT+WS.}
        \label{fig:reconstruction_residuals}
    \end{figure}

    Residuals for the various test cases are shown in Figure \ref{fig:reconstruction_residuals}. It can be seen from the figure that generally, the absolute residual in amplitude gets smaller for increased frequency coverage (Case 1 to 3). When looking at $\chi$, CRAFT methods give smaller residuals than RM Synthesis. Additionally, CRAFT+WS provides similar, if not better, reconstructions of the polarization angle as seen by the similar or smaller residuals for CRAFT+WS.

\end{document}